\documentclass{SCIS2023}
\usepackage{listings}
\usepackage{makecell}
\usepackage{multirow}

\newcommand{\tabincell}[2]{\begin{tabular}{@{}#1@{}}#2\end{tabular}}
\newcommand{\whiteding}[1]{\ding{\numexpr171+#1\relax}}
\newcommand{\bi}{\begin{itemize}}
\newcommand{\ei}{\end{itemize}}
\newcommand{\be}{\begin{enumerate}}
\newcommand{\ee}{\end{enumerate}}

\newcommand{\eg}{{e.g., }}

\newcommand{\etal}{{et al.}}

\def\sysfullname{{cycle-task-oriented memory protection (CToMP) }}
\def\sysname{{CToMP }}
\def\sysnameend{{CToMP}}

\begin{document}
\ArticleType{RESEARCH PAPER}
\Year{2022}
\Month{}
\Vol{}
\No{}
\DOI{}
\ArtNo{}
\ReceiveDate{}
\ReviseDate{}
\AcceptDate{}
\OnlineDate{}

\title{CToMP: A Cycle-task-oriented Memory Protection Scheme for Unmanned Systems}{CToMP: A Cycle-task-oriented Memory Protection Scheme for Unmanned Systems}

\author[1]{Chengyan MA}{}
\author[1]{Ning XI}{}
\author[2]{Di LU}{{dlu@xidian.edu.cn}}
\author[3]{Yebo FENG}{}
\author[1]{Jianfeng MA}{}

\AuthorMark{Ma C Y}

\AuthorCitation{Ma C Y, Xi N, Lu D, et al}


\address[1]{School of Cyber Engineering, Xidian University, Xi'an {\rm 710071}, China}
\address[2]{School of Computer Science and Technology, Xidian University, Xi'an {\rm 710071}, China}
\address[3]{University of Oregon, Eugene {\rm 97403}, USA}

\abstract{Memory corruption attacks (MCAs) refer to malicious behaviors of system intruders that modify the contents of a memory location to disrupt the normal operation of computing systems, causing leakage of sensitive data or perturbations to ongoing processes.
Unlike general-purpose systems, unmanned systems cannot deploy complete security protection schemes, due to their limitations in size, cost and performance. 
MCAs in unmanned systems are particularly difficult to defend against.
Furthermore, MCAs have diverse and unpredictable attack interfaces in unmanned systems, severely impacting digital and physical sectors.

In this paper, we first generalize, model and taxonomize MCAs found in unmanned systems currently, laying the foundation for designing a portable and general defense approach. According to different attack mechanisms, we found that MCAs are mainly categorized into two types---\textit{return2libc} and \textit{return2shellcode}.
To tackle \textit{return2libc} attacks, we model the erratic operation of unmanned systems with cycles and then propose a \sysfullname approach to protect control flows from tampering.
To defend against \textit{return2shellcode} attacks, we introduce a secure process stack with a randomized memory address by leveraging the memory pool to prevent \texttt{Shellcode} from being executed. 
Moreover, we discuss the mechanism by which \sysname resists the ROP attack, a novel variant of \textit{return2libc} attacks.
Finally, we implement \sysname on CUAV V5+ with Ardupilot and Crazyflie. 
The evaluation and security analysis results demonstrate that the proposed approach \sysname is resilient to various MCAs in unmanned systems with low footprints and system overhead.}

\keywords{unmanned system, memory corruption attack, memory protection, system security, randomized memory address}

\maketitle

\section{Introduction}
\label{sec:intro}
Unmanned systems are embedded computing systems that monitor, respond to, or control an external environment through sensors, actuators, and other input/output interfaces~\cite{DBLP:journals/csur/Stankovic96}. 
Such systems must meet various constraints, such as timing, efficiency, security, etc., that are imposed on them by real-time behaviors of the external world they interface with. 
These unmanned systems wide ranging applications, such as search and rescue~\cite{6290694}, agriculture~\cite{rs12091491}, and autonomous vehicles.
However, the increasing popularity of unmanned systems increases security concerns regarding them~\cite{chai2019short,zhi2020security}. According to recent studies~\cite{8538989,8462886,1712371}, current unmanned systems continue to possess a myriad of flaws that can be leveraged by malicious parties to launch attacks, affecting their proper operations, stealing private data of users, or even endangering public safety.

Among all the threats encountered by unmanned systems, software-oriented attacks are emerging and gradually becoming one of the most concerning.
These attacks exploit software vulnerabilities within the unmanned system firmware to maliciously interfere with system operations, however, it has received attention only recently~\cite{10.1007/978-3-319-40667-1_4,DBLP:conf/ndss/NieslerSD21,272210}. 
Particularly, \textbf{memory corruption attacks (MCAs)}~\cite{regalado2015gray}, a special form of software-oriented attacks, are becoming increasingly rampant~\cite{10.1145/2592798.2592824,DBLP:conf/ndss/KimKCGL0X18,272132}. 
Unlike the segmented memory management in general computer systems, the user code in unmanned systems shares the same physical memory with the kernel. 
This design enables the user code to directly access, invoke, or even tamper with critical kernel instructions. Furthermore, unmanned systems usually allow memory access from peripherals (\eg ZigBee and WiFi), thereby allowing malicious parties to launch MCAs even wirelessly. 
For example, in UAV systems, attackers can launch MCAs to modify the return addresses of stack frames remotely, thereby seizing the control of drones through memory overflows.
Recent studies~\cite{7795496,DBLP:conf/ndss/KimKCGL0X18} have demonstrated the feasibility and risk of such MCAs in the real world. Therefore, designing a practical and effective methodology to defend against MCAs is vital for ensuring the reliability and stability of unmanned systems.

\begin{figure}[!t]
\centering
\begin{minipage}[c]{0.48\textwidth}
\centering
\includegraphics[width=\linewidth]{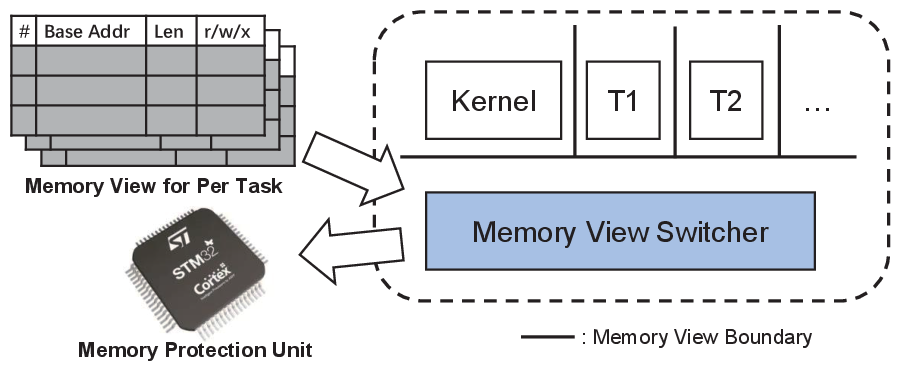}
\end{minipage}
\hspace{0.02\textwidth}
\begin{minipage}[c]{0.48\textwidth}
\centering
\includegraphics[width=\linewidth]{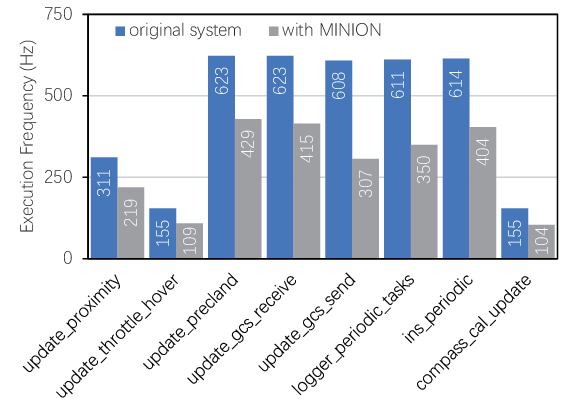}
\end{minipage}\\[3mm]
\begin{minipage}[t]{0.48\textwidth}
\centering
\caption{Architecture of MINION~\cite{DBLP:conf/ndss/KimKCGL0X18} for defending against MCAs. Here, each task has a memory view to indicate its accessible memory regions. }
\label{fig:type1_arch}
\end{minipage}
\hspace{0.02\textwidth}
\begin{minipage}[t]{0.48\textwidth}
\centering
\caption{Effect of MINION on unmanned system performance. }
\label{fig:type1}
\end{minipage}
\end{figure}

Multiple approaches have been proposed to tackle MCAs in unmanned systems. Their principal methodology is equipping unmanned systems with memory management mechanisms to isolate the kernel from the user code. 
Widely used techniques to achieve memory isolation include TrustZone~\cite{9833604} and Memory Protection Unit (MPU)~\cite{10.1145/2592798.2592824,DBLP:conf/ndss/KimKCGL0X18,215981}. 
Because the microcontroller units (MCUs) used by unmanned systems rarely have TrustZone, more approaches focusing on employing MPU are required to tackle MCAs (e.g., \textbf{MINION}~\cite{DBLP:conf/ndss/KimKCGL0X18}). 
According to multitasking and modular programming features in unmanned systems, these approaches first create a memory view for the user code of each task from the perspectives of code reachability, I/O interfaces, and in-memory data. This process can be done manually or semi-automatically using Low-Level Virtual Machine (LLVM)~\cite{1281665}. They then leverage MPU to securely switch memory views in task switchover, as shown in Figure~\ref{fig:type1_arch}. Although this methodology can limit MCAs to a certain extent, it has several drawbacks. First, by applying such approaches, programmers must consider memory range divisions in developments, which will confuse developers. Moreover, existing approaches can cause considerable system overhead, reducing the processing power of unmanned systems. We conducted some performance tests by implementing MINION, the most representative of these approaches, on a UAV based on CUAV V5+ hardware with firmware Ardupilot. We found that some low-priority tasks in MINION cannot reach the execution frequency in the original system because of the time and system overheads caused by the memory view switching and MPU configuration (Figure~\ref{fig:type1}). This flaw is particularly fatal to unmanned systems with strict real-time requirements. For example, the execution frequency of tasks \texttt{update\_gcs\_send} and \texttt{ins\_periodic} is reduced by half. This reduction caused a severe problem: we could not receive the Mavlink messages sent by the drone to the ground station in time, and the drone could not perceive its own acceleration and other states in real time. Eventually, the drone lost control and crashed. Moreover, such approaches only use MPU to protect the memory security of unmanned systems. However, because of the hardware performance limitation, the upper limit of the memory regions MPU can protect is 16; hence, these approaches lack extensibility in the face of complex unmanned systems. 

To fill this gap, herein, we propose an effective and efficient approach called \sysfullname to protect unmanned systems from MCAs. Unlike existing approaches that isolate memory regions for each task, \sysname treats tasks executed within one cycle as a whole and focuses on protecting a few critical codes, variables, and registers. To reduce the system overhead and ensure the timeliness of tasks, \sysname releases the pressure of memory view switching and MPU configuration by performing security operations and memory allocations right before starting each cycle rather than before beginning each task. To enhance the security and prevent the execution of injected \texttt{Shellcode}, \sysname dynamically assigns the process stack address and buffer addresses by randomizing memory allocations. 

Compared with previously reported approaches to resist MCAs, \sysname makes the following contributions:

\bi
    \item To simplify the defense interface, we summarize the existing MCAs for unmanned systems and classify them into two categories (i.e., \emph{return2libc} and \emph{return2shellcode}) based on their execution of \texttt{Shellcode}.
    \item We establish a cycle-based operation model for unmanned systems as the foundation for our system design. To our knowledge, \sysname is a brand-new memory protection approach for unmanned systems. Moreover, we build a secure process stack based on randomized memory allocation to enhance system security. 
    \item By evaluating \sysname on two unmanned platforms (CUAV V5+ with Ardupilot and Crazyflie), its effectiveness and efficiency are verified. Our approach defends against MCAs without compromising the real-time performance of unmanned systems. 
\ei

The remainder of this paper is organized as follows. After describing the security model and motivations in \S\ref{sec:back}, 
we present the design of \sysname in \S\ref{sec:design}. 
Further, we evaluate \sysname in \S\ref{sec:eva}. 
Finally, we describe related work in \S\ref{sec:related} and conclude this paper in \S\ref{sec:con}.

\section{Security Model \& Motivation}
\label{sec:back}
In this section, we elaborate on necessary background knowledge to help readers understand our system design. Later, to simplify the defense interface, we summarize and taxonomize existing MCAs against unmanned systems, classifying them into two categories (i.e., \emph{return2libc} and \emph{return2shellcode}). 
Finally, we demonstrate the motivations of our proposed approach. 

\subsection{Background}
\label{sec:opera}
As for power consumption, hardware platforms of unmanned systems are still dominated by low-cost MCUs, such as STM32 series based on the ARM Cortex-M architecture. 
To ensure the availability of systems, the development of unmanned systems focuses more on realizing more applications with limited software and hardware resources rather than security protection that may take up more resources. 
However, these low-cost MCUs are also designed with security features, which have not been taken seriously. 
We will briefly introduce these features. 

\begin{figure}[t]
\centering
\includegraphics[width=0.7\linewidth]{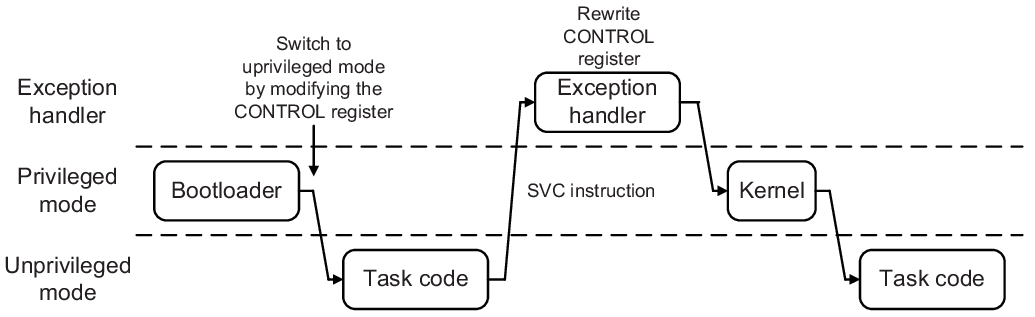}
\caption{Execution level switching in unmanned systems. 
The system generally boots from privileged mode and then converts to unprivileged mode to execute tasks. 
If kernel instructions are required, the system will switch to privileged mode. }    
\label{fig:modeswitch}
\end{figure}

\subsubsection{Execution Levels}
\label{subs:mode}
Privileged and unprivileged modes are two execution levels of MCUs~\cite{PM0253}. 
In privileged mode, the code can invoke all available instructions and access all resources (\eg system registers, memory, and peripherals). 
In unprivileged mode, the code only has limited access to some resources. 
Any access to unpermitted registers, including MPU, SysTick timer, and Nested vectored interrupt controller (NVIC), will raise hardware exceptions. 
Therefore, the kernel can be placed in privileged mode, and user code related to the unmanned system task implementation can be placed in unprivileged mode. 

The \texttt{CONTROL} register determines whether the code executes in privileged or unprivileged mode. 
Only in privileged mode can the code modify the \texttt{CONTROL} register, thereby changing the execution level to unprivileged mode. 
Correspondingly, in unprivileged mode, the code cannot change the execution level by directly modifying the \texttt{CONTROL} register. 
Instead, it must call the \texttt{SVC} instruction to switch the system to a handler mode, in which the code can change to privileged mode, as shown in Figure~\ref{fig:modeswitch}. 

\subsubsection{Stack Pointer}
\label{subs:stack}
\begin{table}[!t]
\footnotesize
  \caption{Summary of execution levels and stack use options. }
  \label{tab:mode}
  \def\tabblank{\hspace*{10mm}}
  \begin{tabularx}{\textwidth}
  {@{\tabblank}@{\extracolsep{\fill}}ccc@{\tabblank}}
    \toprule
    \textbf{Execution level}& \textbf{Used to execute}& \textbf{Stack used}\\
    \midrule
    Privileged& \tabincell{c}{Kernel \\Exception handlers}& Main stack\\\midrule
    Unprivileged& Tasks& Process stack\\
  \bottomrule
\end{tabularx}
\end{table}

Another key point is that the code will use a separate descending stack in privileged and unprivileged mode, respectively.
Consequently, MCUs must implement two stacks, the \emph{main stack} and the \emph{process stack}. 
Meanwhile, each stack has an independent stack pointer holding the address of the last stacked item in memory; we call them the \texttt{MSP} and \texttt{PSP}. When the code needs to invoke kernel operations, it typically switches to the \emph{main stack}, and the tasks always use the \emph{process stack}. 
We summarize these features in Table ~\ref{tab:mode}. 

\subsubsection{Memory Protection Unit}
\label{subs:mpu}
Cortex-M is the most commonly used ARM processor for embedded unmanned systems. 
MPU is a security kernel feature of the Cortex-M series MCUs~\cite{AN5156}. It can set the properties and access permissions of different memory addresses by dividing the memory map into several regions, such as whether a certain address range is allowed to be executed, read, or written. 
Furthermore, MPU can isolate system resources and code by limiting access permissions in privileged and unprivileged modes. If the code accesses a memory region that MPU protects without permission, the processor will throw a fault exception, which can prevent illegal memory access. 

\subsection{Security Model}
\label{sec:security}
\begin{figure*}[!t]
\centering
\includegraphics[width=0.9\textwidth]{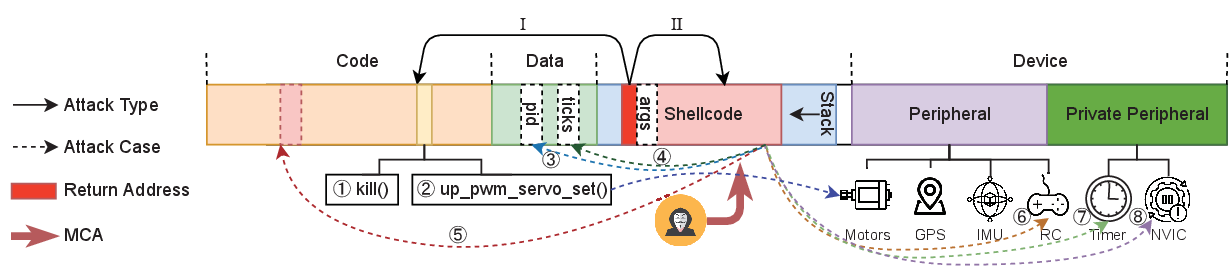}
\caption{MCAs against unmanned systems are generally implemented by executing \texttt{Shellcode}, and their targets are distributed in the code, data and device areas of memory. }
\label{fig:mca}
\end{figure*}

\begin{table}[!t]
  \caption{Attack cases and their attack targets. We divide these attack cases into two types (i.e., \emph{return2libc} and \emph{return2shellcode}) according to the different implementation methods. }
  \label{tab:mac}
  \footnotesize
  \begin{tabularx}{\textwidth}{clcp{8.3cm}}
    \toprule
    \textbf{No.}& \makecell[c]{\textbf{Attack Case}}& \textbf{Attack Type}& \makecell[c]{\textbf{Attack Target}}\\
    \midrule
    \whiteding{1}& Process Termination& \multirow{3}*{return2libc}& Reusing \texttt{kill()} function in the \textbf{Code} area.\\
    
    \whiteding{2}& Servo Operation& & Invoking \texttt{up\_pwm\_servo\_set()} in the \textbf{Code} area with two abnormal args (\texttt{channel} and \texttt{value}) to change the speed of motors.\\\midrule
    
    \whiteding{3}& Control Parameter Attack& \multirow{8}*{return2shellcode}& Overwriting the PID parameters in the \textbf{Data} area.\\

    \whiteding{4}& Soft Timer Attack& & Overwriting \texttt{ticks} and \texttt{last\_run} two count parameters in the \textbf{Data} area.\\

    \whiteding{5}& Memory Remapping& & Copying the malicious code to \textbf{FLASH} and replacing the existing function in the \textbf{Code} area.\\
    
    \whiteding{6}& RC Disturbance& & Modifying RC-related registers in the \textbf{Peripheral} area.\\
    
    \whiteding{7}& Hard Timer Attack& & Reloading the system timer value of \texttt{SYST\_RVR} in the \textbf{Private Peripheral} area.\\
    
    \whiteding{8}& Interrupt Vector Overriding & & Overriding \texttt{NVIC} registers in the \textbf{Private Peripheral} area.\\
  \bottomrule
\end{tabularx}
\end{table}

In MCAs attackers use memory vulnerabilities to inject \texttt{Shellcode} into the stack to achieve various malicious operations. 
As shown in Figure~\ref{fig:mca}, \texttt{Shellcode} can be executed in two ways. 
Type \uppercase\expandafter{\romannumeral1} involves overwriting the return address in the stack to the existing function address in memory so that the device can perform some operations that are not allowed or illogical; it is called \emph{return2libc}. 
Type \uppercase\expandafter{\romannumeral2} involves overwriting the return address in the stack with the starting address of \texttt{Shellcode} to execute the malicious behavior customized by the attacker. We name it \emph{return2shellcode}. 
Additionally, MCAs for embedded unmanned systems are subdivided into several cases in Table~\ref{tab:mac} according to different \texttt{Shellcode} attack targets. 
Notably, as a variant of \emph{return2libc} attack, the ROP attack~\cite{10.1145/1315245.1315313, 10.1145/2133375.2133377} has been widely discussed on x86 architecture systems. This attack executes instruction sequences, called \emph{gadgets}, in the existing code area through the stack overflow vulnerability, and each gadget ends with a \texttt{ret} instruction. By chaining these gadgets together with \texttt{Shellcode}, an attacker can modify registers or some memory data. In the unmanned system scenario, the ROP attack is similar to attack case \whiteding{2} in Table~\ref{tab:mac}.

Because of hardware and performance limitations, the existing embedded development platforms do not have memory isolation or address randomization and other security protection measures that can effectively avoid MCAs. 
Therefore, we believe that MCAs are ubiquitous in embedded unmanned systems. 
We assume the attackers have the following abilities to find and exploit memory vulnerabilities to perform attacks. 

\begin{itemize}
\item Attackers can use Fuzzing~\cite{DBLP:journals/iee/MouzaraniSZ16,5954459,7371847} and other testing methods to discover memory vulnerabilities, such as stack overflow in the communication protocol encoding/decoding of the system remote control. 
\item Attackers can use static analysis tools ~\cite{9152796,10.1145/2897845.2897900,10.1145/3296957.3177157,10.1145/3432893} to reverse the firmware in embedded hardware and obtain the memory addresses of sensitive parameters, such as the PID parameters. 
\item Attackers can discover which key registers are accessed during the sensor reading and writing process through firmware Re-Hosting~\cite{10.1145/3423167,247646,242028,272151}.
\end{itemize}

Because \texttt{Shellcode} must be injected from outside the system, the vulnerabilities that can be easily exploited by attackers lie in the user code interacting with the outside rather than in the kernel responsible for interrupt management, task scheduling and other internal features. 
After discovering the vulnerabilities, attackers write \texttt{ShellCode} and tamper with the return address of the stack frame to achieve MCAs. 
Because all codes in an MCU share the same physical memory, an attacker who launches an attack from the user code can modify key parameters in the memory by injecting malicious instructions. 
In addition to attackers, ordinary users may also modify key parameters in the kernel because of misoperation, such as setting unreasonable PID parameters through MAVLink commands~\cite{10.1145/3510003.3510084}, which can also have disastrous consequences.

\subsection{Finding of Drawbacks in Relevant Works}
\label{sec:motiv}
At present, the most effective solution to MCAs is to implement memory isolation between the kernel space and user code, which limits the range that malicious code can access in memory~\cite{10.1145/2592798.2592824,DBLP:conf/ndss/KimKCGL0X18,10.1145/3400302.3415727,215981}. 
These relevant works achieve memory isolation by building a memory view for each task, just like running a task in a \textbf{Sandbox}, and we summarize them as \textbf{task-oriented} solutions. Task-oriented memory isolation requires configuring MPU registers for the next task after one task ends, which is called memory view switching. 
Since MPU registers are accessible only in privileged mode, memory view switching must enter an exception handler and switch mode while the task runs in unprivileged mode. We summarize two stages in memory view switching as follows: 

First, before task application execution, the following steps are required: 
\begin{itemize}
\itemindent 2.8em
\item[(1)] The scheduler reads the task information list and decides which task must be executed next. 
\item[(2)] The system initializes the process stack required by the task code and switches the memory view to configure the memory regions that the task can access by MPU. 
\item[(3)] The system switches from the main stack to the process stack. 
\item[(4)] The system switches the execution level from privileged mode to unprivileged mode. 
\end{itemize}

Second, to switch from unprivileged mode to privileged mode after task completion, the following steps are required: 
\begin{itemize}
\itemindent 2.8em
\item[(1)] The system enters \texttt{SVC} interrupt through a system call, and switches from unprivileged mode to privileged mode. 
\item[(2)] Then, the system switches the stack pointer from \texttt{PSP} to \texttt{MSP}.
\end{itemize}

\begin{table}[!t]
  \caption{Time cost of each step in the execution level and memory view switching. Since each task must configure different numbers of accessible memory regions, the time spent by MPU also differs. }
  \label{tab:timecost}
  \footnotesize
  \def\tabblank{\hspace*{10mm}}
  \begin{tabularx}{\textwidth}
  {@{\tabblank}@{\extracolsep{\fill}}c|cc@{\tabblank}}
    \toprule
    \textbf{Stage} & MPU Configure& Stack Initialize and Switch\\
    \midrule
    \textbf{Time} ($\mu$sec)& 9-15& 10\\\midrule
    \textbf{Stage} &SVC System Call& Execution Level Switch\\\midrule
    \textbf{Time} ($\mu$sec)& 1& 1\\
  \bottomrule
\end{tabularx}
\end{table}

Through experiments, we also sorted out the time spent in each step, as shown in Table ~\ref{tab:timecost}. 
However, when we applied this task-oriented solution to \emph{Ardupilot}, an open-source unmanned system firmware, efficiency problems emerged. 
In the search for the cause, we found that although the task scheduling of unmanned systems is uncertain and irregular in the traditional concept, we can use a cycle-based model to describe it. 
Unmanned systems perform a certain number of tasks during each cycle. We will describe this model in detail in \S\ref{sec:cycle}. 
Since at least seven tasks need executing in one cycle in Ardupilot, it takes at least 154$\mu$sec for frequent memory view switching between tasks. Consequently, some low-priority tasks cannot obtain the time in a cycle to execute, which affects unmanned system availability. 
Therefore, we consider whether we can use MPU to manage the memory access range of a cycle, which is similar to putting all tasks in a cycle into a sandbox rather than putting one task into one sandbox. This approach reduces the frequency of memory view switching, thereby improving system efficiency. 

However, we need to prevent attackers from attacking tasks in the same cycle. When we simultaneously manage several tasks in a cycle with MPU, tasks do not securely access the code and data from each other because they share the same region of memory. 
We determine the static loading of the code as the reason why embedded unmanned systems are vulnerable to MCAs. This attribute allows attackers to easily analyze the call stack and memory addresses and inject \texttt{Shellcode} to cause damage. 
Therefore, we hope to implement random memory allocation, which is not widely used in embedded unmanned systems because of the lack of the Memory Management Unit (MMU), and improve the difficulty of memory analysis, so as to achieve the purpose of protecting user code and data in the same cycle. 

\section{Design}
\label{sec:design}
In this section, we elaborate on the design of \sysnameend.
According to the use characteristics and requirements of unmanned systems, we have determined the following design goals: 
\be
    \item[\textbf{G1}] \textbf{Security. }Our design needs to be resistant to MCAs. 
    \item[\textbf{G2}] \textbf{Efficiency. }After our design is added to unmanned systems, 
    it only brings a minimum runtime overhead, 
    and cannot affect the real-time requirement of task execution. 
    \item[\textbf{G3}] \textbf{Extensibility. }Our design can adapt to different functional requirements 
    in different scenarios rather than being fixed. 
    \item[\textbf{G4}] \textbf{Low Footprint. }Due to the limitation of memory size in unmanned systems, 
    our design cannot occupy too much memory space. 
    \item[\textbf{G5}] \textbf{Generality. }Our design can be easily adapted to different unmanned devices 
    without much additional development work. 
\ee
Following these design goals, 
We first model the operation of unmanned systems in \S\ref{sec:cycle}.
Then, we introduce the overall system architecture in \S\ref{sec:sysarch}. 
In addition, we describe the structure of the secure process stack in \S\ref{sec:psp} and the workflow of \sysname in \S\ref{sec:flow}. 

\subsection{Cycle-based Model of Unmanned Systems}
\label{sec:cycle}
\begin{table}
  \caption{RTOSes used by five kinds of unmanned system firmware, and their application scenarios. }
  \label{tab:multitask}
  \footnotesize
  \def\tabblank{\hspace*{10mm}}
  \begin{tabularx}{\textwidth}
  {@{\tabblank}@{\extracolsep{\fill}}cccc@{\tabblank}}
    \toprule
    \textbf{Firmware}& \textbf{RTOS}& \textbf{\tabincell{c}{Number\\ of Tasks}}& \textbf{Application Scenarios}\\
    \midrule
    Ardupilot& ChibiOS& 49& \makecell[l]{UAVs, Rovers, Submarines, ...}\\
    PX4& Nuttx& 27& \makecell[l]{Drones, VTOLs, Rovers, ...}\\
    FMT& RT-Thread& 6& \makecell[l]{UAVs, Cars, Robots, ...}\\
    Paparazzi& ChibiOS& 9& \makecell[l]{Rotorcrafts, Hybrids, Boats, ...}\\
    Crazyflie& FreeRTOS & 37& \makecell[l]{Drones}\\
  \bottomrule
\end{tabularx}
\end{table}

\begin{figure}[!t]
\centering
\begin{minipage}[c]{0.42\textwidth}
\centering
\includegraphics[width=\linewidth]{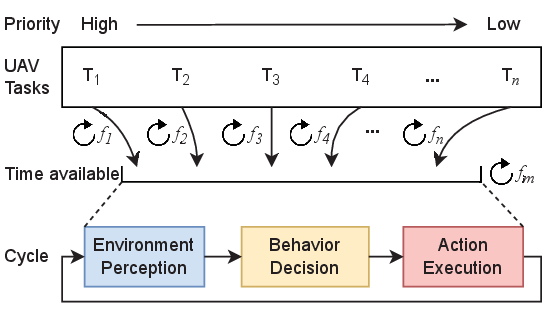}
\end{minipage}
\hspace{0.02\textwidth}
\begin{minipage}[c]{0.54\textwidth}
\centering
\includegraphics[width=\linewidth]{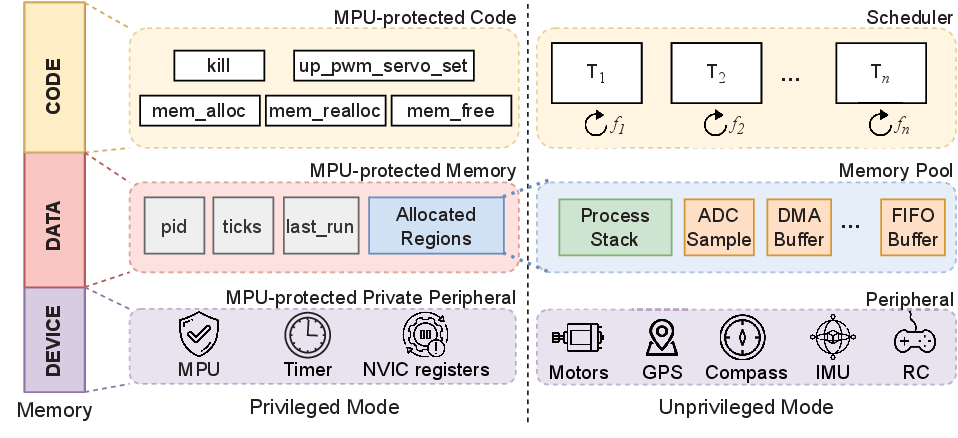}
\end{minipage}\\[3mm]
\begin{minipage}[t]{0.42\textwidth}
\centering
\caption{Cycle-based operation model of unmanned systems. }
\label{fig:multitask}
\end{minipage}
\hspace{0.02\textwidth}
\begin{minipage}[t]{0.54\textwidth}
\centering
\caption{Overall architecture of \sysnameend. }
\label{fig:sysarch}
\end{minipage}
\end{figure}

Unmanned systems encapsulate simple functions such as sensor reading/writing and actuator execution into multiple tasks, and realize more complex remote control or autonomous operations through the cooperation between tasks. 
In the traditional concept, the unmanned system, as a real-time system, requires each task to be completed at a certain time, and these tasks are triggered uncertainly by the external environment. 
For example, we cannot predict when the unmanned system will encounter an obstacle and change its trajectory. 
Table~\ref{tab:multitask} summarizes the results of our research on various unmanned system firmware applicable to drones, rovers, and boats. 
However, we found that although the number of tasks in these firmware varies, they all rely on the real-time operating system (RTOS) to implement task scheduling. 
Further, the scheduling algorithms in these RTOSes are composed of preemptive scheduling and round robin. 
This means, in practical design, unmanned systems only need to implement three stages of functions---\textbf{environment perception}, \textbf{behavior decision} and \textbf{action execution}, which are supported by tasks within a specified time period to meet the real-time requirement, and we call this time period a cycle. 
As shown in Figure~\ref{fig:multitask}, we regularize seemingly uncertain and irregular task executions in unmanned systems with cycle-based modeling, thereby laying the foundation for our system design. 
In detail, $f_m$ in Figure~\ref{fig:multitask} represents the cycle frequency, that is, how many cycles will be experienced in one second, and the longest available duration of one cycle is $1/f_m$ seconds. Then for task $i$, its frequency $f_i$ is determined by the average cycle interval (ACI) at which it can be scheduled:
\begin{equation}
f_i=f_m/ACI_i
\label{equ:ic}
\end{equation}
For example, in an unmanned system with a cycle frequency of 400Hz, a task is executed every two cycles on average, then the frequency of this task is 200Hz.
What's more, each task is given a priority. 
In a cycle, high-priority tasks must be scheduled first, while low-priority tasks will use the remaining time as needed after the execution of high-priority tasks. 
This enables each task to meet its functional requirements as much as possible. 
Therefore, it can be obtained that the less time a cycle takes, the faster the cycle frequency becomes, and each task can be scheduled in a more timely manner, so the real-time requirement of the unmanned system is guaranteed. Conversely, if the single cycle time is long, the system reliability will be reduced.

\subsection{System Architecture}
\label{sec:sysarch}
The main goal of our system design is to prevent attackers from exploiting victim tasks with memory vulnerabilities to access kernel code or modify memory data. 
Specifically, our approach should be able to defend against each attack case in \S\ref{sec:security} effectively. As illustrated in Figure~\ref{fig:sysarch}, in our system architecture, the code segment, data segment, and device address segment in the memory are separated by MPU into two environments: privileged mode and unprivileged mode. According to the assumptions in \S\ref{sec:security}, as user code for all tasks is executed in unprivileged mode, the \texttt{Shellcode} injected by an attacker to exploit memory vulnerabilities usually occurs in unprivileged mode. 
Unlike relevant task-oriented approaches that require designing a memory view for each task and attack to defend against MCAs, our solution focuses on blocking two attack paths of \texttt{Shellcode} to secure the unmanned systems. 
We explain how this architecture meets the design goal \textbf{G1} \emph{Security} as follows. 

\noindent\textbf{Defense against \emph{return2libc}:} Since functions located in privileged mode cannot be directly called by user code running in unprivileged mode, for attack cases \whiteding{1} and \whiteding{2}, we place two functions \texttt{kill} and \texttt{up\_pwm\_servo\_set} in privileged mode. 
The \texttt{kill} is a kernel function used by RTOS in unmanned systems to terminate the task. 
The \texttt{up\_pwm\_servo\_set} is a function used to control the execution of actuators in unmanned systems. 
Malicious calls to both of these functions can have catastrophic consequences, such as the crash of drones and changing the driving path of unmanned vehicles. Therefore we configure them with MPU not to be accessible in unprivileged mode to prevent \texttt{Shellcode} written by attackers from maliciously jumping to their function addresses. 

\noindent\textbf{Defense against \textbf{ROP}:} According to the description in \S\ref{sec:security}, the ROP attack achieves the goal of executing malicious code by exploiting gadgets of existing code and concatenating them. But as we explained in defending against \emph{return2libc} attacks, the code containing sensitive parameters has been protected by MPU, so attackers cannot jump to gadgets that can modify these parameters through \texttt{Shellcode}. Although attackers can still exploit gadgets located in unprivileged mode, these gadgets cannot access registers and data protected by MPU; hence, our architecture can still effectively resist the ROP attack.

\noindent\textbf{Defense against \emph{return2shellcode}:} In this attack way, the attacker can execute more malicious instructions in \texttt{Shellcode} and destroy more memory data. 
Since all code in the MCU shares the same physical memory, in the absence of memory isolation, the attacker's targets include, but are not limited to, the PID parameters that control the unmanned system attitude, the count parameters responsible for task scheduling, and various peripherals such as sensors and system Timer. 
Here we use two solutions to protect them. 

First, for attack cases \whiteding{3}, \whiteding{7}, and \whiteding{8}, the data and registers in these attack objects should be configured and fixed in memory when the unmanned system is initialized. 
Therefore, we only need to configure these parameters and registers to be unmodifiable in unprivileged mode with MPU after they are set. They can be effectively protected from being accessed by the attacker's \texttt{Shellcode}. At the same time, we configure the entire code segment to be non-writable with MPU, so that attack case \whiteding{5} cannot overwrite the existing code with malicious instructions through \texttt{Shellcode}.

For attack case \whiteding{4}, \texttt{ticks} and \texttt{last\_run}, these two count parameters are used for task scheduling, and they are changed in each cycle to ensure that the execution of each task can meet the functional requirements. Parameter \texttt{ticks} is the number of cycles that have passed, \texttt{last\_run} is an array used to calculate the interval between the last executed cycle of each task and the current cycle. 
Therefore, we set the two parameters to be modifiable only in privileged mode. When the tasks of a cycle are executed, the system switches the execution level once, and then updates these two parameters after entering privileged mode, which can satisfy the task scheduling and parameter security. 

Finally, for some code and data that must share the memory with other tasks, such as reading remote control variables from registers that can only run in unprivileged mode (attack case \whiteding{6}), we design another solution to protect their security. 
By analyzing the execution condition of \emph{return2shellcode}, we found that the attacker needs to obtain the starting address of injected \texttt{Shellcode} to jump to the malicious code execution area by tampering with the return address in the stack. 
Due to the static loading code of MCUs, it is easy for an attacker to analyze the required memory address and implement malicious behavior. 
However, we noticed that using the two instructions \texttt{\_\_set\_PSP} and \texttt{\_\_set\_CONTROL(SP\_PROCESS)} can specify an area in memory as the process stack used by tasks. If we randomize the stack in each cycle, it will be much more difficult for an attacker to analyze the starting address of \texttt{Shellcode}, which can effectively prevent the execution of \texttt{Shellcode}. 
Therefore, we design a \textbf{memory pool} in unprivileged mode to dynamically allocate a random address area for the stack used in each cycle. 

In summary, \sysname can resist all attack cases and meet the design goal \textbf{G1} \emph{Security}. 
At the same time, different from task-oriented solutions, we manage the memory access range of tasks in a cycle as a whole, so that the MPU configuration and execution level switching do not need to be performed between tasks, but only before the start of a cycle, so as to satisfy the design goal \textbf{G2} \emph{Efficiency}. 
In addition, since MPU in MCUs can protect up to 16 memory regions, it is difficult for solutions that only rely on MPU to play a protective role when the functions of unmanned systems are gradually complex. 
Our solution can effectively prevent the execution of \texttt{Shellcode} by randomizing the process stack, and is not limited by the number of MPU-protected memory regions and other hardware resources, which meets the design goal \textbf{G3} \emph{Extensibility}. 
What's more, we only add a memory pool area into the system architecture to achieve dynamic memory allocation, and we will describe how the memory pool satisfies the design goal \textbf{G4} \emph{Low Footprint} in \S\ref{sec:psp}. 
Last but not least, MPU, execution levels and the process stack are features supported by most unmanned system MCUs, so our system architecture can be easily adapted to other unmanned system firmware and RTOSes, and our design realizes the \textbf{G5} \emph{Generality}. 

\subsection{Secure Process Stack}
Unfortunately, due to the lack of MMU support, MCUs for unmanned systems do not support dynamic memory allocation, or can only allocate memory blocks with fixed addresses, which does not meet our needs for randomizing the process stack. 
For example, only three fixed regions are provided for memory allocation to choose from in ChibiOS, which supports the unmanned system firmware Ardupilot. 
In order to be compatible with the existing memory allocation solutions in unmanned systems, we designate an area in memory as a memory pool.
We use the following structure array $R=\{r_0,r_1,\dots,r_n\}$ to denote the usage of the memory pool. 

\begin{lstlisting}[language=c++]
typedef struct allocated_region {
    void *pointer;
    uint32_t start_address;
    uint32_t size;
}allocated_region[MAX_ALLOCATE_NUM];
\end{lstlisting}

In this structure, we define a pointer to the allocated memory region, the start addresses of the region and the allocation size. At the same time, we also need to specify the maximum number of memory regions that can be allocated by defining the value of \texttt{MAX\_ALLOCATE\_NUM}. 
Three functions are designed to allocate, re-allocate and free memory:

\begin{lstlisting}[language=c++]
void *mem_alloc(uint32_t size);
void *mem_realloc(uint32_t size);
void mem_free(void *pointer);
\end{lstlisting}

\begin{algorithm}[t]\footnotesize
\caption{Allocation and release of memory regions for the secure process stack. 
Here, $r_i$ is a unit in the structure array $R$. }
    \begin{algorithmic}[1]
    \renewcommand{\algorithmiccomment}[1]{#1}
    \LOOP [\textsc{MemAlloc}($size$)]
        \IF {$allocated\_regions\_num>$\texttt{MAX\_ALLOCATE\_NUM}}
            \RETURN NULL
        \ENDIF
        \STATE $start\_addr \gets TRNG()$
	    \STATE $end\_addr \gets start\_addr + size - 1$
	    \FOR {$i = 0 \to allocated\_regions\_num$}
	        \IF {MAX($r_i.start\_addr, start\_addr$) $<$ MIN($r_i.start\_addr+r_i.size-1, end\_addr$)}
	            \RETURN mem\_realloc(size)
	        \ENDIF
	    \ENDFOR
	    \STATE $R \gets R ~\cup$ New Region
	    \STATE $allocated\_regions\_num++$
	    \RETURN $start\_addr$
	\ENDLOOP
    \LOOP [\textsc{MemRealloc}($size$)]
        \renewcommand{\algorithmiccomment}[1]{\hfill $\triangleright$ #1}
        \IF[Retry times can be set.] {$realloc\_times>3$}
            \RETURN NULL
        \ELSE
            \RETURN mem\_alloc(size)
        \ENDIF
    \ENDLOOP
	\LOOP [\textsc{MemFree}($pointer$)]
	    \FOR {$i = 0 \to allocated\_regions\_num$}
	        \IF{$r_i.pointer = pointer$}
	            \STATE Delete $r_i$ from $R$
	            \STATE $allocated\_regions\_num--$
	            \RETURN
	        \ENDIF
	    \ENDFOR
	\ENDLOOP
	\end{algorithmic}
\label{alog:memalloc}
\end{algorithm}

As described in Algorithm~\ref{alog:memalloc}, \texttt{mem\_alloc} uses the True Random Number Generator (TRNG)~\cite{RM0410} supported by ARM Cortex-M series MCUs to generate a start address of the memory area to be allocated. 
Then, when the number of allocated regions does not reach the maximum setting, the allocator needs to traverse all memory regions in the structure array \texttt{allocated\_region} to check whether there is a conflict between the to-be-allocated region and the allocated regions. 
If there is no conflict, the allocator should save the region information into the structure array. 
Otherwise, it will find the allocatable region again by \texttt{mem\_realloc}. 
To free the allocated memory, it just needs to call \texttt{mem\_free} to delete the information of the pointer in the structure array. 

\begin{table}[!t]
  \caption{The five memory regions that need to be dynamically allocated in the unmanned system firmware Ardupilot, and their sizes. }
  \label{tab:buffer}
  \footnotesize
  \def\tabblank{\hspace*{10mm}}
  \begin{tabularx}{\textwidth}
  {@{\tabblank}@{\extracolsep{\fill}}c|ccc@{\tabblank}}
    \toprule
    \textbf{Name} & ADC sample& DMA buffer& RX bounce buffer\\
    \midrule
    \textbf{Size} (Byte)& 144& 304& 64\\
    \midrule
    \textbf{Name} & TX bounce buffer& FIFO buffer& Process stack\\
    \midrule
    \textbf{Size} (Byte)& 64& 112& 1024\\
  \bottomrule
\end{tabularx}
\end{table}

In order to measure the space occupancy of the memory pool, we summarize the stack and buffers that use memory allocation and the size of regions they require in Table~\ref{tab:buffer}. 
It can be seen only five buffers and one stack need to dynamically allocate memory; the value of \texttt{MAX\_ALLOCATE\_NUM} should be set to 6, which means that the space occupied by the structure array \texttt{allocated\_region} is 72 bytes. 
At the same time, the total memory space required for these buffers and the process stack is 1712 bytes. 
In order to ensure the randomness of the process stack space address, we set the size of the memory pool to 5632 bytes, which is about three times the size of the total required space. 
Since most MCUs have 1MB-2MB of Flash ROM and 192KB-512KB of SRAM, the memory pool we designed has a very small footprint and can meet the design goal \textbf{G4}. 

\label{sec:psp}
\begin{figure*}[t]
\centering
\includegraphics[width=0.9\textwidth]{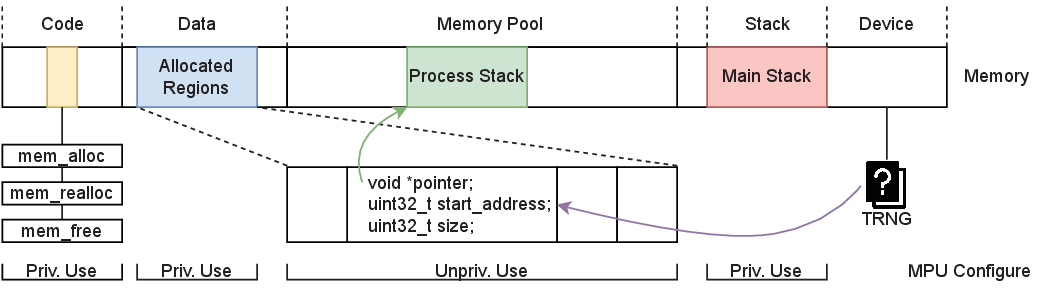}
\caption{Secure process stack with randomised addresses. We used MPU to configure the permissions of each module in randomized memory allocation. }    
\label{fig:psp}
\end{figure*}

On the other hand, in order to secure the memory allocation, we need to configure the three functions and the struct array \texttt{allocated\_region} to be accessible only in privileged mode with MPU. 
Only the memory pool can be accessed by tasks in unprivileged mode. 
Due to the real-time requirement of unmanned systems, the data in these buffers that need to dynamically allocate memory are only valid for one cycle; that is, they will not be available in the next cycle. 
So we allocate memory for them at the beginning of a cycle and release them at the end of the cycle, which can effectively meet their usage requirements. 
This also means that we do not need to perform frequent execution level switching in order to securely allocate memory by MPU. 
As illustrated in Figure~\ref{fig:psp}, we implement a secure process stack to resist \texttt{Shellcode} execution. 

\subsection{Workflow of \sysnameend}
\label{sec:flow}
\begin{figure}[t]
\centering
\includegraphics[width=0.6\linewidth]{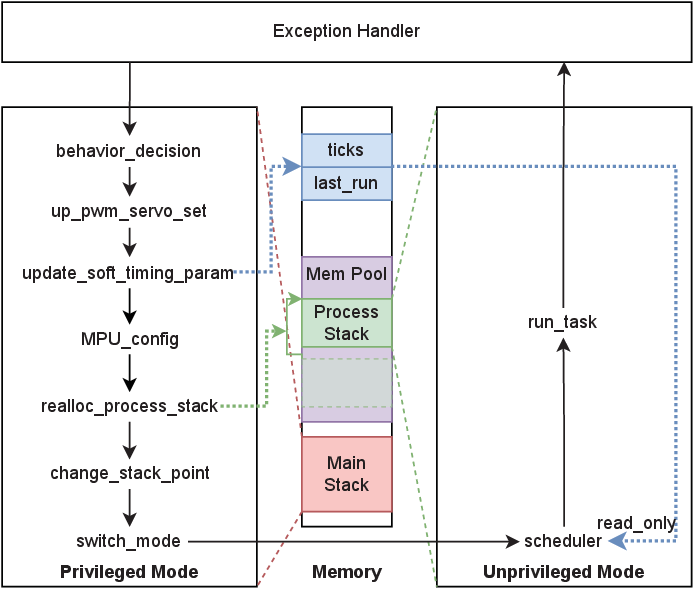}
\caption{\sysname exection flow. } 
\label{fig:flow}
\end{figure}

Our approach works upon cycle-based task modeling, where three stages of functions (i.e., environment perception, behavior decision, and action execution in one cycle) are embedded within. The workflow of \sysname corresponds to the three stages of functions in one cycle.

First, use MPU to configure hardware and software resources that can be accessed in one cycle. 
Then, a region is allocated in the memory pool for tasks of this cycle as a secure process stack. At the same time, we also need to allocate regions in the memory pool for other buffers. 
Next, we use the \texttt{\_\_set\_CONTROL} instruction to make the stack pointer point to the secure process stack and switch the execution level to unprivileged mode. 

In unprivileged mode, each task is scheduled on demand based on the priority and two soft timing parameters (\texttt{ticks} and \texttt{last\_run}) that are read-only in this mode. 
At this stage, the unmanned system uses various sensors and communication equipment to complete the perception of the surrounding environment and its own state and receive control commands. 

After tasks in a cycle are executed, the unmanned system needs to switch back to privileged mode to call some key functions protected by MPU, implement behavior decision and action execution, and allocate memory resources for the next cycle. 
Therefore, the user code is required to call the \texttt{SVC} instruction into the exception handler. 
In exception handling, switch the execution level back to privileged mode, and replace the running stack with the main stack after exiting the exception. 
In addition, we can also update soft timing parameters in this mode. 
We describe the above workflow with Figure~\ref{fig:flow}. 

According to the unmanned system model in \S\ref{sec:cycle}, we conclude that one of the ways to ensure the real-time performance of unmanned systems is to shorten the running time of a cycle. 
Therefore, we clarify the advantages of \sysname in the design goal \textbf{G2} \emph{Efficiency} by analyzing the computation cost of \sysname in one cycle and comparing it with the task-oriented approach. 
We assume that $n$ tasks need to be executed in one cycle, and the task-oriented memory protection method uses MPU to limit $m_i$ accessible areas for each task to prevent damage attacks caused by memory leaks. 
At the same time, since the tasks are all executed in unprivileged mode, and MPU can only be configured in privileged mode, each time a task is executed, it needs to go through the mode switch twice to configure MPU and stack for the next task before return to unprivileged mode. 
Therefore, in a task-oriented manner, the computation cost in one cycle is: 
\begin{equation}
Time_{task-oriented}=\sum_{i=1}^{n}m_iT_{MPU}+n(2T_{switch}+T_{Stack}+T_{SVC})
\label{equ:to}
\end{equation}
In Eq.~(\ref{equ:to}), $T_{MPU}$ denotes the time cost for MPU to configure a memory area, $T_{Stack}$ denotes the time cost of the process stack initialization, $T_{SVC}$ is the time consumed in one SVC call, and $T_{switch}$ is the time cost of one mode switch. These indicators' values can be found in Table~\ref{tab:timecost}. 

From the workflow of \sysname in Figure~\ref{fig:flow}, it can be seen that since we regard all tasks in one cycle as a whole, \sysname mode only switches twice in one cycle. That is, switching to the unprivileged mode when entering a new cycle. After the cycle ends, switching to privileged mode to process sensitive data with an SVC call. What's more, due to only protecting key code and data, the memory regions configured by MPU pre-cycle are also fewer. We suppose these memory regions are $z$ blocks.
The computation cost of \sysname is as follows: 
\begin{equation}
Time_{\sysname}=zT_{MPU}+2T_{switch}+T_{Stack}+T_{SVC}
\label{equ:our}
\end{equation}
By comparing Eq.~(\ref{equ:to}) and~(\ref{equ:our}), we can obtain that the computation cost of \sysname in one cycle is much smaller than that of the task-oriented scheme, so \sysname meets design goal \textbf{G2}. More discussion on performance evaluation will be detailed in \S\ref{sec:p_eva}.

\section{Evaluation}
\label{sec:eva}
In this section, we first introduce the implementation and configuration details. We then evaluate the proposed approach from the following perspectives: 
\begin{itemize}
\item How effectively \sysname can compete against memory corruption attacks? 
\item What is the performance impact of \sysname on the unmanned system?
\end{itemize}

\subsection{Implementation \& Configuration}
\begin{figure}[t]
\centering
\includegraphics[width=0.9\linewidth]{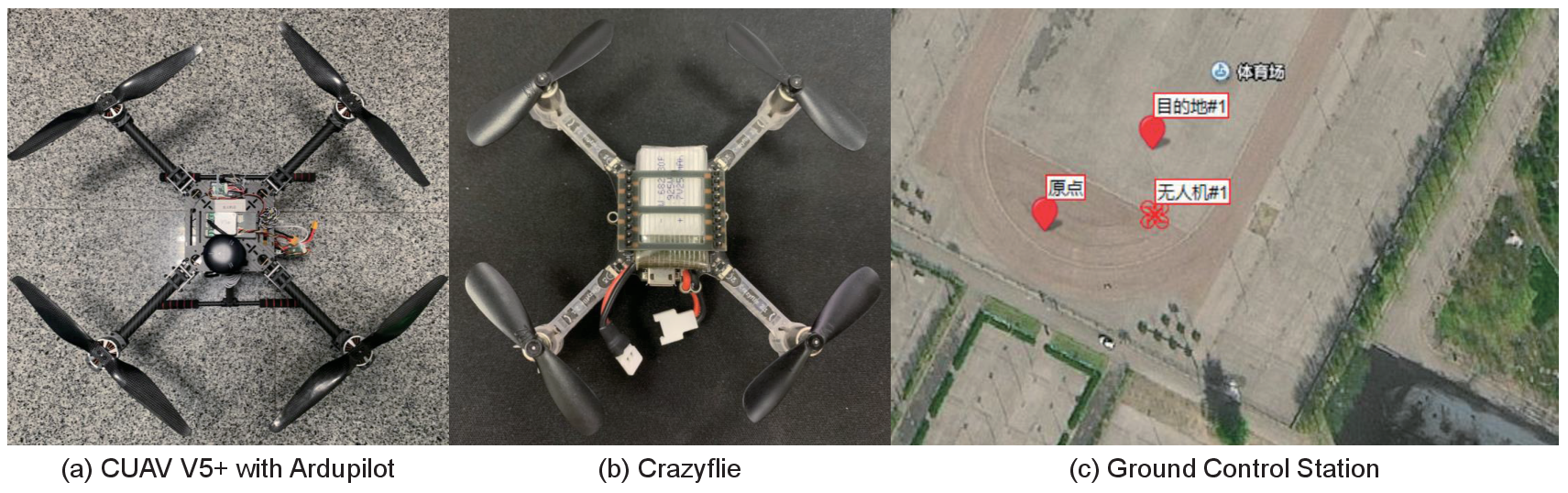}
\caption{Experimental platforms and our ground control station. (a) CUAV V5+ with Ardupilot; (b) Crazyflie; (c) Ground Control Station. }
\label{fig:exper}
\end{figure}

We implement \sysname on CUAV V5+, a typical unmanned aerial vehicle that is full compatibility with the Pixhawk project FMUv5 design standard. 
CUAV V5+ is equipped with an ARM 32-bit Cortex-M7 processor, a 512KB SRAM and a 2MB Flash memory where the code and data are stored. 
Similar to most unmanned systems designed based on the Pixhawk standard, CUAV V5+ supports the open source firmware, Ardupilot, which uses ChibiOS as the real-time operating system. 
We also chose an open source drone, \emph{Crazyflie}, which is equipped with an ARM 32-bit Cortex-M4 CPU, a 196KB SRAM and a 1MB Flash memory and uses FreeRTOS as the operating system. 
We use Crazyflie as a supplementary experiment to verify the generality of our solution. 
Figure~\ref{fig:exper} shows our experiment platforms and a simple ground control station (GCS) to record the flight status of our UAVs so that we can analyze the experimental results. 

\subsection{Study Case: CUAV V5+ with Ardupilot}
\subsubsection{Security Analysis}
Since unmanned systems are closely related to the physical world, attacks against them tend to cause damage and loss in the real world. 
Similarly, MCAs in Ardupilot mainly target the key code and data stored in the memory that can affect the stable flight of drones~\cite{8462886}. 
In the following, we describe several representative attacks in detail. 

\noindent\textbf{Process Termination. }ChibiOS, the real-time operating system in Ardupilot, provides some POSIX-like interfaces. The \texttt{kill} function can directly terminate the execution of tasks. 
We modified the return address in the stack used by the vulnerable user code, and terminated \texttt{fast\_loop}, the core task of Ardupilot, by calling \texttt{kill} through \emph{return2libc}. 
However, under the protection of \sysname, \texttt{kill} cannot be called outside of privileged mode, which makes it impossible for attackers to use the process stack in unprivileged mode to perform malicious jumps. 

\noindent\textbf{Servo Operation. } The function of \texttt{up\_pwm\_servo\_set} is to output the PWM waveform to control the motor speed of drones. Similar to the previous attack, this attack changed the return address to \texttt{up\_pwm\_servo\_set} address, and passed two illegal parameters (\texttt{channel} and \texttt{value}) to this function, which caused motors rotating abnormally. 
Same as the previous one, \sysname makes \texttt{up\_pwm\_servo\_set} inaccessible in unprivileged mode, thus preventing the attack. 

\begin{figure}[t]
\centering
\subfloat[Roll angle values under control parameter attack]{\includegraphics[width=.35\linewidth]{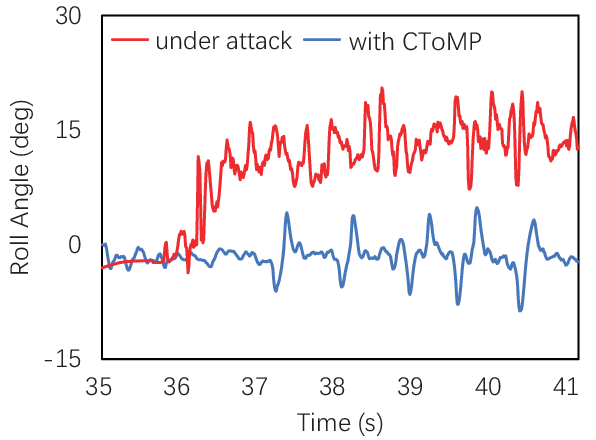}\label{fig:roll}}
\hfil
\subfloat[Pitch angle values under control parameter attack]{\includegraphics[width=.35\linewidth]{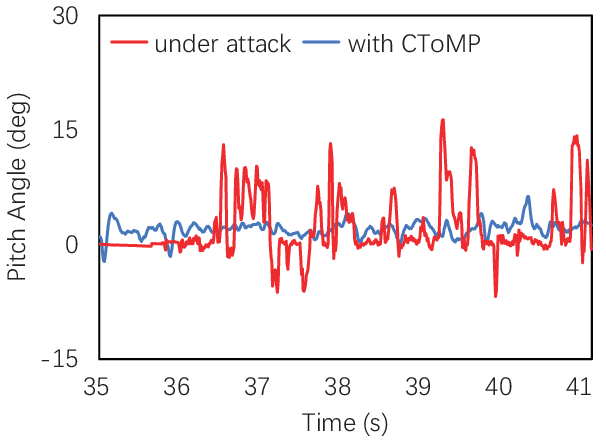}}
\hfil
\subfloat[Yaw angle values under control parameter attack]{\includegraphics[width=.35\linewidth]{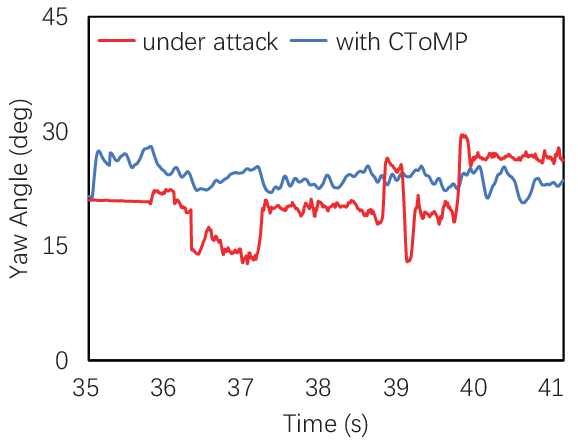}\label{fig:yaw}}
\hfil
\subfloat[Altitude values under soft timer Attack]{\includegraphics[width=.35\linewidth]{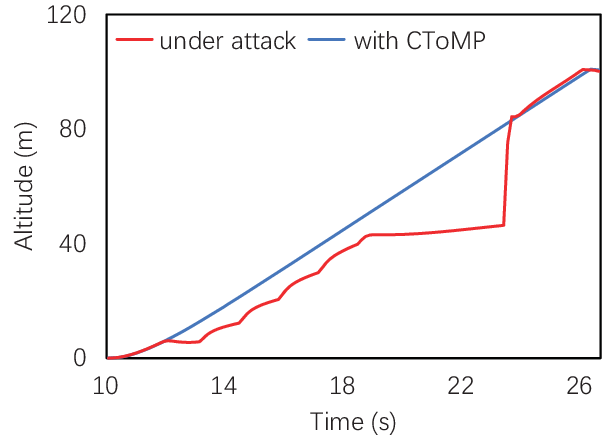}\label{fig:alt}}
\hfil
\caption{Under different attacks, protected system sensor output compared to the unprotected system.}
\label{fig:evasecurity}
\end{figure}

\noindent\textbf{Control Parameter Attack. }PID controller is the most widely used automatic controller. It has the advantages of simple principle, easy implementation, and wide application. 
There are three important parameters \texttt{pid\_rate\_roll}, \texttt{pid\_rate\_pitch}, \texttt{pid\_rate\_yaw} in Ardupilot, which respectively control the roll, pitch and yaw angle of the drone aircraft attitude. 
After the aircraft debugging is completed, these three parameters will be fixed and written into the memory and will not change; slight changes in parameters will affect the stability of flight. 
In our experiment, the attack overwrote the PID control parameters by injecting malicious code while the drone was hovering in the air. The drone began to sway back and forth, left and right, gradually deviating from the original position. 
In our architecture, we configure these three parameters to be \textbf{read-only} using MPU, protecting them from tampering, then the attack will no longer work. 
Figure~\ref{fig:roll} - Figure~\ref{fig:yaw} record the data of three angles output by IMU. It is obvious that under the protection of \sysname, the output of IMU is more stable. 

\noindent\textbf{Soft Timer Attack. }The scheduler is an important component of unmanned systems, and it is responsible for the on-demand execution of various functional tasks. We interfered with the execution frequency of task \texttt{update\_altitude} by modifying the two soft timer parameters \texttt{ticks} and \texttt{last\_run} in the scheduler. 
Figure~\ref{fig:alt} shows the flight altitude recorded by the GCS in both cases when the drone is attacked and protected by \sysname. 
It can be seen when an attack occurs, although we increased the remote control (RC) throttle to make the drone start rising, the change of altitude is not reflected on the GCS in real-time, which undoubtedly interferes with the driver's control of the drone, bringing a certain risk. 
We effectively avoid this attack by placing scheduling-related data in privileged mode that malicious code cannot access. 

\noindent\textbf{Memory Remapping. }ARM Cortex-M series MCUs allow users to perform a hot-patching operation. 
This means the new program can be written into the Flash ROM, not through \texttt{JTAG} or \texttt{USART0}, but through \texttt{USB}, \texttt{RS232}, wireless transmission and other interfaces that are still preserved in the external environment after hardware packaging. 
The attacker can replace the existing function in Flash with the malicious code in \texttt{Shellcode} from the vulnerable user code, and the work of the program update should only be the responsibility of \textbf{Bootloader}. 
Therefore, in our architecture, after Bootloader starts the system, the entire code segment is configured as unwritable by MPU, making this attack ineffective. 

\noindent\textbf{RC Disturbance. }Most tasks of unmanned systems are to sense the surrounding environment and receive control commands. In order to isolate from the kernel, we put the user code of these tasks into unprivileged mode. 
For the cause of facilitating the management of these sensors and communication devices, ARM Cortex-M series MCUs provide \textbf{MMIO} technology, which can map the registers of these devices to the peripheral area in memory. 
This allows tasks to mutually access data from other devices, which facilitates sensor data fusion, but attackers can exploit vulnerable code to attack some sensors across tasks. 
For example, by modifying the channel value of RC registers in \texttt{Shellcode}, the attacker can even take control of the unmanned system. 
In our architecture, due to the support of the randomized process stack, it is very difficult for an attacker to find the starting address of \texttt{Shellcode} injected through the vulnerable code, and cannot perform malicious operations. 

\noindent\textbf{Hard Timer Attack. }All ARM Cortex-M series MCUs have a 24-bit system timer, \textbf{SysTick}. This timer counts down from the reload value (\texttt{SYST\_RVR}) to zero, providing the unmanned system with a microsecond-accurate system clock. 
Through this clock, the unmanned system can count the execution time of tasks in each cycle and schedule tasks efficiently. 
In the absence of memory isolation, an attacker can slow down the system time by overwriting the value of \texttt{SYST\_RVR}, which will seriously affect the task scheduling of unmanned systems. 
Since SysTick is a private peripheral in MCUs, its \texttt{SYST\_RVR} can only be reloaded in privileged mode. 
In our architecture, vulnerable user code in unprivileged mode cannot be exploited to attack the hard timer. 

\noindent\textbf{Interrupt Vector Overriding. }The \textbf{NVIC} supports 1 to 240 interrupts for each task in unmanned systems. 
Tasks can be configured with a priority in the range of 0-255, and this information is stored in a vector table. 
By replacing the priorities in the original vector table and increasing the priority of some tasks, the attack can make some unimportant tasks always be called in each cycle, which will inevitably waste software and hardware resources. 
However, like the hard timer attack, the NVIC is also a private peripheral, so this attack cannot be implemented in our architecture. 

\subsubsection{Performance Evaluation}\label{sec:p_eva}
\begin{figure*}[t]
\centering
\includegraphics[width=\textwidth]{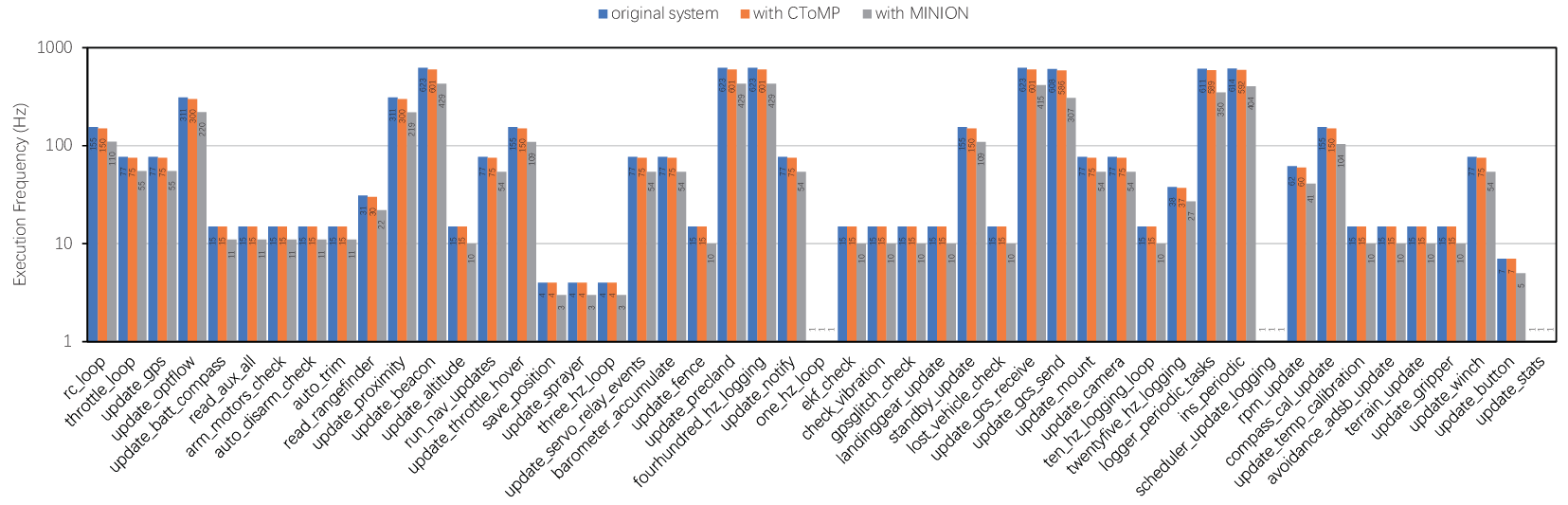}
\caption{Log-based the execution frequency of tasks with \sysname and with MINION and baseline in original system Ardupilot. }
\label{fig:evaalltasks}
\end{figure*}

In order to measure the impact of new security features on the real-time performance of unmanned systems, some relevant works in recent years are based on the time of task execution. 
They believe that as long as tasks are completed within the maximum executable time, it will not affect the real-time performance of unmanned systems. 
However, according to our analysis of some existing unmanned system firmware, we found that the initial-setup maximum executable time of these tasks is inaccurate, and they will change dynamically during the operation of unmanned systems to improve the scheduling efficiency. 
This is why we found that the relevant works have problems with some complex unmanned systems, such as Ardupilot. 
The maximum executable time is not an accurate measure of whether the functionality of the task is affected. 

According to the operation model of unmanned systems in \S\ref{sec:opera}, we believe that a task needs to achieve a certain execution frequency to meet its functional requirement. 
Failure to reach the execution frequency will affect the real-time performance of unmanned systems, such as perception delay and command delay, which will directly affect the safety of unmanned systems. 
Therefore, we choose task execution frequency as an indicator to measure the performance impact of security features on unmanned systems. 
As demonstrated in Figure~\ref{fig:evaalltasks}, we can see that our approach brings little system overhead. None of the task execution frequency is affected under the protection of \sysname, enabling the protected system to reach the same operation efficiency as the original system
In the framework of relevant work \textbf{MINION}, many low-priority tasks are affected by the frequent configuration of MPU and cannot be executed in a cycle, and the desired execution frequency cannot be achieved. 
Therefore, the effect of \sysname on the efficiency of Ardupilot is completely acceptable. 

\begin{figure}[t]
\centering
\subfloat[From small to large. ]{\includegraphics[width=.45\linewidth]{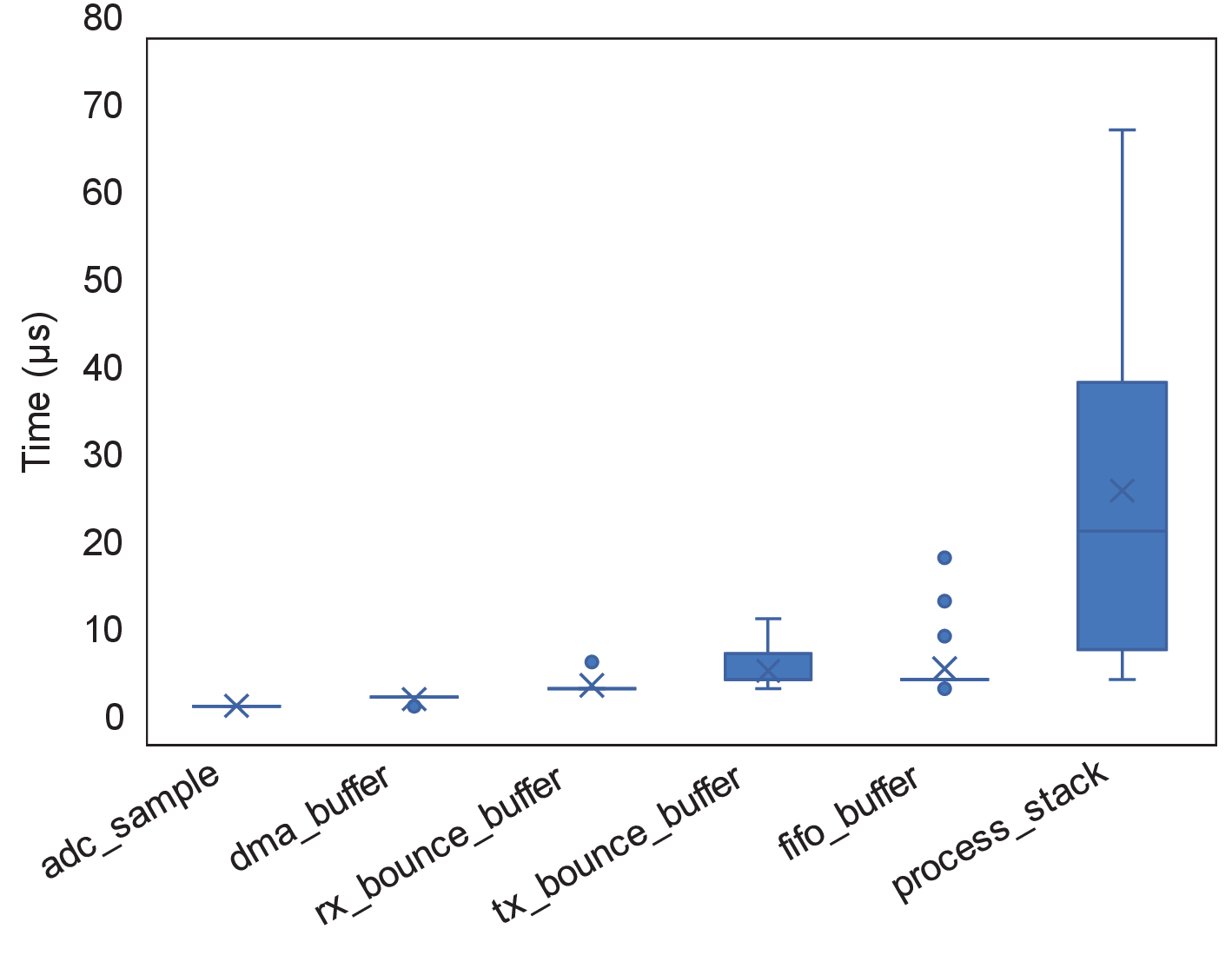}\label{fig:mem1}}
\hfil
\subfloat[From large to small. ]{\includegraphics[width=.45\linewidth]{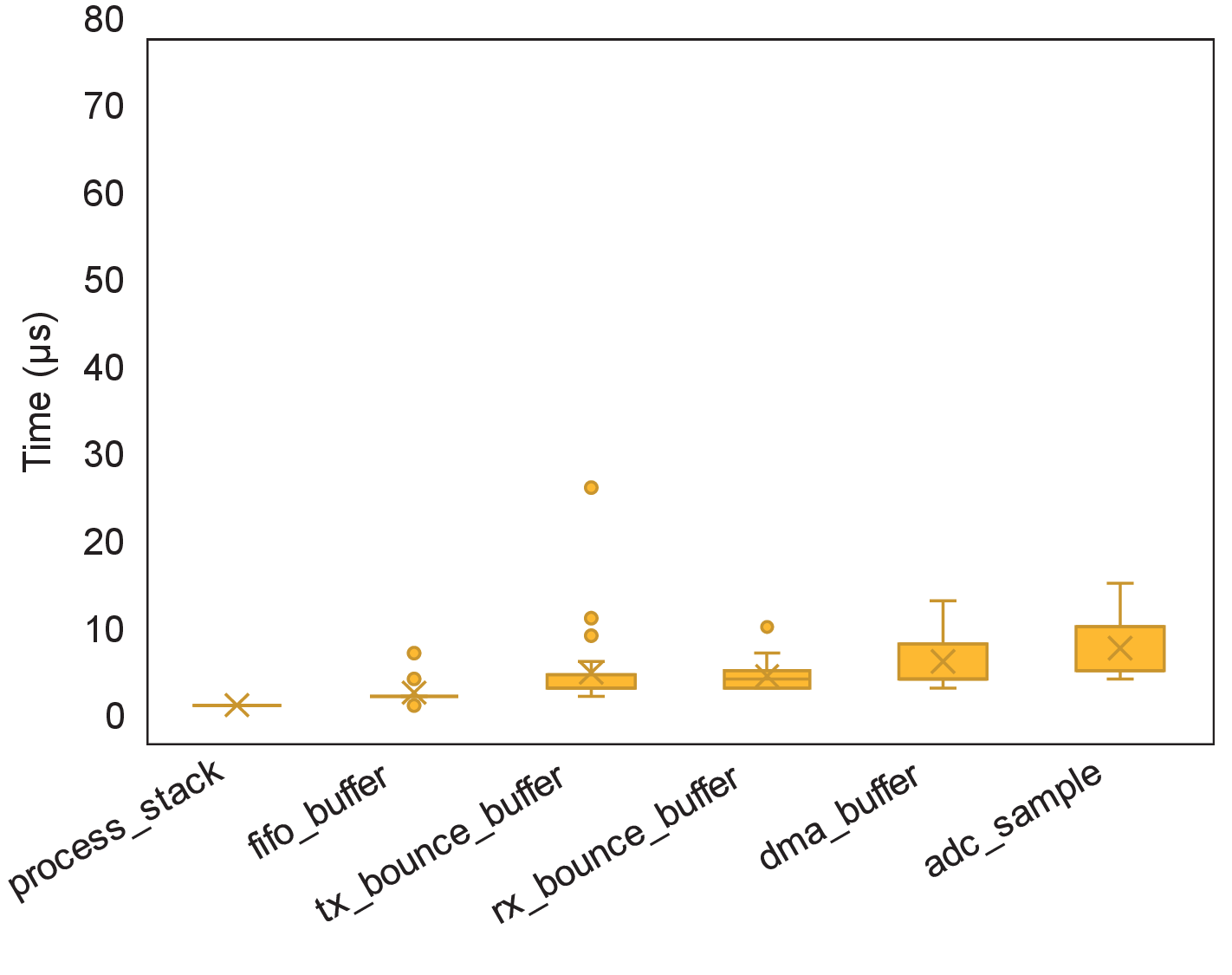}\label{fig:mem2}}
\caption{Runtime Overhead of two kinds of memory allocation solutions. }
\label{fig:evamem}
\end{figure}

At the same time, we noticed that randomizing memory allocation can lead to memory fragmentation issues. This causes a large number of address conflicts in memory allocation, and multiple memory reallocations will consume more time. We tried the simplest solution, which is to arrange the memory regions that need to be allocated in order of size. We found that allocating memory from large to small (Figure~\ref{fig:mem2}) can improve the efficiency by 26.5\% than allocating memory from small to large (Figure~\ref{fig:mem1}). 
The time of memory allocation also can be accepted. 

\subsection{Study Case: Crazyflie}
\label{app:crazy}
\begin{figure}[!t]
\centering
\begin{minipage}[c]{0.3\textwidth}
\centering
\includegraphics[width=\linewidth]{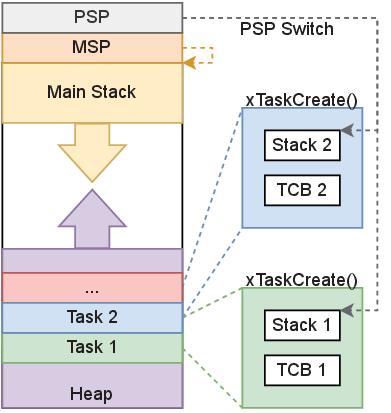}
\end{minipage}
\hspace{0.02\textwidth}
\begin{minipage}[c]{0.55\textwidth}
\centering
\includegraphics[width=\linewidth]{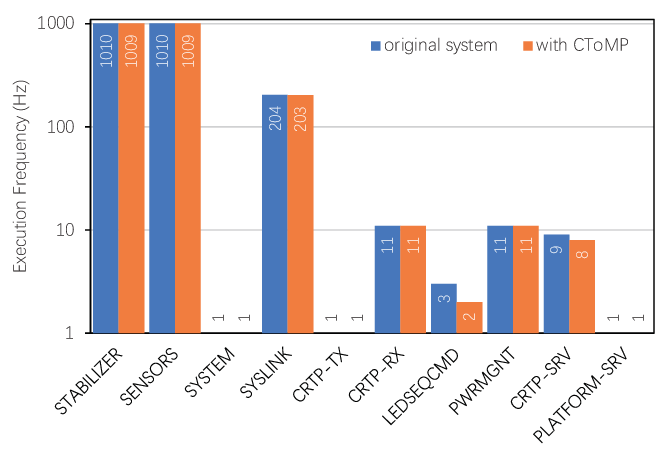}
\end{minipage}\\[3mm]
\begin{minipage}[t]{0.3\textwidth}
\centering
\caption{Task Control Block (TCB) in FreeRTOS. }
\label{fig:freertos}
\end{minipage}
\hspace{0.02\textwidth}
\begin{minipage}[t]{0.56\textwidth}
\centering
\caption{Log-based the execution frequency of tasks with \sysname and baseline in original system Crazyflie. }
\label{fig:evacrazy}
\end{minipage}
\end{figure}

To illustrate the generality of \sysname, we validate it on another unmanned system, Crazyflie. 
Since Crzayflie uses the relatively low-end STM32F405 as the MCU, it needs more efficient security functions. 
There are two main missions in Crazyflie, \texttt{STABILIZER} and \texttt{SENSORS}. 
\texttt{STABILIZER} is mainly responsible for the attitude control of drones, while \texttt{SENSORS} is used to collect data from sensors such as gyroscopes and accelerometers. 
At the same time, there is also a Crazy RealTime Protocol (CRTP) service for data communication with the control side. 

Similar to Ardupilot, attackers can exploit vulnerabilities\footnote{https://forum.bitcraze.io/viewtopic.php?t=2063}\footnote{https://forum.bitcraze.io/viewtopic.php?t=4923} in the firmware to inject \texttt{Shellcode}. 
Therefore, in \sysname, we use MPU to put \texttt{STABILIZER} into privileged mode, which is isolated from other user codes that can interact with the outside world. 
This can effectively protect the attitude control of drones from being interfered with the malicious code. 
Unlike Ardupilot, where multiple tasks can use the same process stack, the operating system FreeRTOS in Crazyflie uses \texttt{xTaskCreate} to allocate a fixed stack for each task in RAM, and makes the PSP point to the corresponding stack when the task is scheduled, as illustrated in Figure~\ref{fig:freertos}. 
This eliminates the need to build a memory pool, but only needs to randomly change the stack address of each task in RAM in each cycle. 
We also tested the effect of \sysname on the real-time performance of Crazyflie, as shown in Figure~\ref{fig:evacrazy}. 
In the experiments, the execution frequency of Crazyflie tasks was not affected by \sysname. 

\section{Related Work}
\label{sec:related}
Since the unmanned system is a typical embedded system, the security research of embedded systems also inspires our research. 

\noindent\textbf{MCAs in Embedded Systems. }Although the existing technology strictly checks the integrity of the embedded software, monitoring methods for executing embedded system programs are not widely used because of resource constraints~\cite{8392631}; hence, some attacks can cause memory corruption during runtime, such as stack/buffer overflow attacks and code reuse attacks~\cite{7464886}. 
In particular, memory overflow attacks have been one of the most mainstream methods to date~\cite{8766434,10.2139/ssrn.3429857,8536378}, because in the most commonly used embedded program language C$\backslash$C++, some functions dealing with buffer data lack a boundary detection mechanism, e.g., \texttt{strcpy()}, or because of unavoidable programmer negligence. 
These attacks have always been difficult to solve, and several vulnerabilities related to memory overflow attacks are reported in \textbf{CVE} every day. 
These vulnerabilities are distributed in the firmware of embedded devices that have been released, such as routers of various brands (TP-LINK\footnote{https://cve.mitre.org/cgi-bin/cvename.cgi?name=CVE-2021-44632}\footnote{https://cve.mitre.org/cgi-bin/cvename.cgi?name=CVE-2022-25074}, NETGEAR\footnote{https://cve.mitre.org/cgi-bin/cvename.cgi?name=CVE-2021-45524}), webcams\footnote{https://cve.mitre.org/cgi-bin/cvename.cgi?name=CVE-2021-33549}, and even in the ARM official dependency library, which contains an unsafe function \texttt{encode\_ise()}\footnote{https://cve.mitre.org/cgi-bin/cvename.cgi?name=CVE-2021-44331}. 
These vulnerabilities allow unauthenticated attackers to remotely execute arbitrary code, causing severe losses. 
Furthermore, although few vulnerability reports are available for the UAV real-time system, after analyzing 596 bugs submitted on \emph{github.com} in two types of open-source flight control software, Ardupilot and PX4, Wang \etal~\cite{10.1145/3468264.3468559} found that hackers can exploit some bugs to launch security attacks. 
In Hooper and Tian's research~\cite{7795496}, a buffer overflow bug was used to force a small commercial drone \emph{Parrot} to land without cracking the Wi-Fi password used for control communication. 
Consequently, designing a general memory protection scheme is necessary in embedded real-time systems. 

\noindent\textbf{Memory Protection in Embedded Systems. }To resist the damage caused by MCAs, the most effective approach is to establish a memory management mechanism to partition the memory usage region of the embedded system. 
Some works~\cite{10.1145/1509288.1509295,Bukkapatnam_2020,4959468,7509439,10.1007/978-3-642-35473-1_74} use MMU in general-purpose computing systems for reference to dividing the embedded system memory into blocks and realize dynamic memory allocation. 
However, these schemes have considerable limitations in actual engineering applications. 
For example, \texttt{malloc} function is also implemented in Ardupilot, but it only provides three fixed regions for memory allocation. 

In addition, there are some frameworks~\cite{272132,DBLP:conf/ndss/KimKCGL0X18,10.1145/2592798.2592824,10.1007/978-3-319-40667-1_4} that implement access control to embedded system memory from the perspective of authority management. 
MINION~\cite{DBLP:conf/ndss/KimKCGL0X18} and M2MON~\cite{272132} are two memory protection architectures based on STM32 series chips and are applied to Ardupilot. 
They use MPU to limit the memory-accessible range of tasks' user code that may be hacked by unauthorized users and protect sensitive data from malicious modification through memory corruption vulnerabilities. 
However, with the update of unmanned systems, the functions supported by drones gradually increased, and the performance of these task-oriented solutions declined. In some other low-cost platforms, 
such as Arduino Yun based on the AVR architecture, Sergio \etal~\cite{10.1007/978-3-319-40667-1_4} proposed to use an XOR-based encryption and a liner PRNG to protect the confidentiality of private metadata. 
However, this work is not universal and representative because of the platform specificity of this solution. 
Additionally, we also noticed that to prevent code-reuse attacks (\emph{return2libc} attacks), \cite{10.1007/978-3-319-40667-1_4} and \cite{8937667} used Address Space Layout Randomization (ASLR) technology. However, their approaches can only randomize the code address once during firmware burning or program startup. 
In contrast, \sysname randomizes the stack address when each cycle begins so that the stack address will be different in each cycle, considerably increasing the difficulty of analyzing memory corruption vulnerabilities for attackers. Hence, it prevents \emph{return2shellcode} attacks.

\noindent\textbf{ARM TrustZone Security. }The initial stage of TrustZone~\cite{PRD29} is a security architecture proposed for high-performance Cortex-A processors. 
Recently, low-power Cortex-M33 series MCUs have also begun to support TrustZone~\cite{100688_0200_en}, and STM32L5~\cite{RM0438} is an earlier chip to cover this feature, but it is not widely used. 
However, similar to the switching of execution levels in our system architecture, TrustZone technology in Cortex-M33 divides the secure world and normal world through memory mapping and uses an exception handler to achieve transitions~\cite{10.1145/2660267.2660350,10.1145/3291047}. 
In the future, we can adapt \sysname into the secure world of TrustZone to complete the management of memory resources in the trusted environment. 

\section{Conclusion}
\label{sec:con}
With the widespread use of unmanned systems, the security issues they conceal have gradually gained people's attention. In this paper, we seek to tackle memory corruption attacks (MCAs), which inject malicious code through memory vulnerabilities and tamper with critical kernel instructions or data in memory. 

To achieve this goal, we propose a \sysfullname approach. By analyzing and testing with various typical attack interfaces, we found that \sysname is resilient to different types of MCAs, and it will not affect the efficiency of unmanned systems. To summarize, \sysname is an efficient and dependable memory protection mechanism that can meet the requirements of unmanned systems for velocity, practicality, and reliability simultaneously.
Our source code is available on GitHub: \url{https://github.com/xidian-uav/uav_memory_isolation}.

\Acknowledgements{This work was supported by the National Natural Science Foundation of China (Key Program 62232013), the Fundamental Research Funds for the Central Universities (Nos. ZYTS23202 and YJSJ23007), and the Major Research Plan of the National Natural Science Foundation of China (Grant No. 92267204). We also thank anonymous reviewers and editors for their comments and guidance.}





\end{document}